\documentclass{article}

\usepackage{cite}
\usepackage{amsmath,amssymb,amsfonts}
\usepackage{algorithmic}
\usepackage{graphicx}

\usepackage{subfig}
\usepackage{textcomp}
\usepackage[acronym]{glossaries}
\usepackage{url}
\usepackage{multirow}
\usepackage{soul}
\usepackage{color}
\usepackage[a4paper, total={6in, 8in}]{geometry}
\graphicspath{{./}}

\def\BibTeX{{\rm B\kern-.05em{\sc i\kern-.025em b}\kern-.08em
    T\kern-.1667em\lower.7ex\hbox{E}\kern-.125emX}}

\begin{document}
	
\newacronym{ips}{IPS}{indoor positioning system}
\newacronym{rf}{RF}{radio-frequency}
\newacronym{ble}{BLE}{Bluetooth low energy}
\newacronym{imu}{IMU}{inertial measurement unit}
\newacronym{rfm}{RFM}{reference feature map}
\newacronym{pdr}{PDR}{pedestrian dead reconking}
\newacronym{nls}{NLS}{non-linear least squares}
\newacronym{ins}{INS}{inertial navigation system}
\newacronym{ecdf}{ECDF}{empirical cumulative distribution function}
\newacronym{cep}{CEP}{circular error probable}

\newcommand{\bluehl}[1]{\textcolor{blue}{#1}}
\newcommand{\todo}{\bluehl{\textit{TODO}}}
\newcommand{\tocite}{\bluehl{CITE}~}
\newcommand{\secref}{Sec.~}
\newcommand{\figref}{Fig.~}
\newcommand{\etal}{et~al.~}
\newcommand{\eg}{e.g.,~}
\newcommand{\tabref}{Table~}
\newcommand{\ie}{i.e.,~}


\newcommand{\rev}[1]{\textcolor{black}{{#1}}}
\newcommand{\revtd}[1]{\textcolor{black}{{#1}}}
\newcommand{\revxy}[1]{\textcolor{black}{{#1}}}

\title{Mining geometric constraints from crowd-sourced radio signals and its application to indoor positioning}
\author{\uppercase{Caifa Zhou}, \uppercase{Zhi Li}, \uppercase{Dandan Zeng}, \uppercase{Yongliang Wang}\thanks{Riemann Lab, 2012 Laboratories, Huawei Technologies Co., Ltd, Xi'an, China (e-mail: \{zhoucaifa, lizhi110, zengdandan4, wangyongliang775\}@huawei.com). Corresponding author: Caifa Zhou (e-mail: zhoucaifa@huawei.com).}}

\begin{abstract}
Crowd-sourcing has become a promising \revxy{way to build} a feature-based indoor positioning system \revxy{that has lower labour and time costs}. \revxy{It} can make full use of the widely deployed infrastructure as well as built-in sensors on mobile devices. One of the key challenges is to \revxy{generate} the reference feature map \rev{(RFM)}, a database used for localization, by \rev{aligning} crowd-sourced \rev{trajectories} according to associations embodied in the data. In order to facilitate the data fusion using crowd-sourced inertial sensors and radio signals, \revxy{this paper proposes} an approach \revxy{to} adaptively mining geometric information. \rev{This is the essential for generating spatial associations between trajectories when employing graph-based optimization} \revxy{methods}. The core idea is to estimate the functional relationship \revxy{to map} the similarity/dissimilarity between radio signals to the physical space \revxy{based on} the relative \rev{positions} obtained from inertial sensors and their associated radio signals. Namely, it is adaptable to different modalities of data and \rev{can be implemented in a self-supervised way}. \revxy{We verify the generality of the proposed approach through} comprehensive experimental analysis: i) qualitatively comparing the estimation of geometric mapping models and the alignment of crowd-sourced trajectories\revxy{;} ii) quantitatively evaluating the positioning performance. \rev{The 68\% of the positioning error is less than 4.7 $\mathrm{m}$ using crowd-sourced RFM, which is on a par with manually collected RFM, in a multi-storey shopping mall, which covers more than 10, 000 $ \mathrm{m}^2 $. }

\textbf{Keywords: geometric constraint,
	sensor fusion,
	crowd-sourcing, 
	graph optimization, 
	feature-based indoor positioning}
\end{abstract}



\maketitle

\section{Introduction}
\label{sec:introduction}
{B}{enefiting} from the emergence and popularity of mobile devices and wireless communication networks, \revxy{indoor positioning services with low-cost and sufficient accuracy become} possible. \revxy{Mobile devices play} the role of receivers, \revxy{and can} observe various measurements (e.g., GNSS, IMU data, signals from Wi-Fi/BLE beacons) \revxy{to indicate} the location-relevant information of \revxy{pedestrians}. \revxy{Wireless communication networks provide} a large coverage of location-relevant signals in the environment without additional expense. \revxy{One} promising way \revxy{to build} such an \acrfull{ips} is per crowd-sourcing \revxy{which can obtain} data without the explicit participation of pedestrians. In this way, it can make full use of the large number of in-use mobile devices and the existence of abundant location-relevant signals. Through the crowd-sourcing process, the goal is to construct a \acrfull{rfm}, which includes the location and \revxy{its} corresponding location-relevant features, and can be \revxy{applied} to determine the location of \revxy{pedestrians} \revxy{in} the online \revxy{phase}.

In the case of passively crowd-sourcing relevant data \revxy{to build} an \acrshort{ips}, the main challenge is how to \revxy{integrate} collected data from \revxy{multiple contributors} at different times, \rev{which is equivalent to simultaneously} \revxy{localizing} mobile devices and \revxy{generating} a (feature) map of the environment (\ie SLAM)\cite{liu2019collaborative}. \revxy{The} process of \rev{RFM generation} has to tackle the problem of multi-modalities originated from devices (\ie hardware), \revxy{pedestrian motions}, and signal sources\cite{zhu2020indoor,wang2016indoor}. Among previous proposals\cite{wu2014smartphones,gu2017,gu2019,li2018automatic,yang2013freeloc,zuo2018indoor,nowicki2015indoor,nowicki2019multi,tan2018optimization,li2021wifi}, one promising approach \revxy{to} achieving \rev{data} fusion is trajectory alignment \footnote{\rev{Herein trajectory alignment denotes the estimation of transformation matrix between associated trajectories. This term is borrowed from} \cite{gu2019}.} \rev{via graph-based optimization by identifying} data associations \rev{(i.e., loop closures)} \revxy{contained} in the crowd-sourced data.

When employing \rev{graph-based} optimization for aligning multi-trajectories obtained per crowd-souring, it consists two ends: i) a front-end \revxy{used to construct} associations, which represent the spatial-temporal connectivity naturally embedded in measurable signals; and ii) a back-end \revxy{to find} the optimal alignment between multi-trajectories via minimizing a given loss function. Several well-known open-source software, such as $\text{g}^{2}\text{o}$\cite{grisetti2011g2o} and Ceres \cite{ceres-solver}, can be \revxy{utilized} to perform the optimization. However, there is lack of a general approach to building data associations according to observations. The construction of spatial-temporal associations using \acrfull{rf} measurements is more difficult than applying visual \rev{or range} observations (\eg camera, LiDAR)\cite{taketomi2017visual,cadena2016past}, because there is no well-defined model that can be used to constrain the geometric relationship between \acrshort{rf} measurements. \rev{This paper focuses on automatically mining data associations, which denote geometric constraints between crowd-sourced traces.} 

In order to mitigate the challenge of fusing crowd-sourced data, we propose an approach \revxy{to} constructing the spatial constraints depending only on \acrshort{imu} values and other location-relevant signals (mainly from Wi-Fi and BLE signals). Both measurements are available in crowd-sourced data. For a given similarity measure between \acrshort{rf} features, \revxy{we estimate} a mapping function \revxy{to indicate} the relationship between the similarity in feature space and the geometric distance in geographical space. The mapping function can be used to approximate both the distance between a pair measured radio signals and the level of \revxy{uncertainty} of the distance estimation. Both of them are then used to build the spatial \rev{associations} between trajectories. The proposed approach is to fuse multi-trajectories per \rev{graph-based} optimization in order to align \rev{relatively} located trajectories. It yields a \acrshort{rfm}, which can be \revxy{applied} to feature-based indoor positioning, \rev{with low-cost}.

\rev{We summarize the main contributions of this paper as follows:}
\begin{enumerate}
	\item \rev{We propose an approach to adaptively modelling the geometric information embodied in radio signals in order to characterize the uncertainty of data associations in a self-supervised manner.}
	\item \rev{We present a framework that integrates crowd-sourced location-relevant features and dead reckoning information from various mobile devices for feature-based indoor positioning in unknown environments with large scale.}
	\item \rev{We evaluate our approach in multi-storey shopping malls with an area over 10,000 $ \mathrm{m}^2 $ (per floor) in the way of commercialized crowd-sourcing, which does not require any interaction from data contributors.}
\end{enumerate}

In the following sections, we first give a brief overview on previous work regarding \acrshort{rfm} construction and geometric constraints generation in the \revxy{field of} feature-based indoor positioning. This is then followed by a formal definition of the problem of information fusion formulated as \rev{graph-based} optimization. In \secref\ref{sec:spa_rf} \revxy{we present the approach to} adaptively modelling the geometric information using the relative locations and radio signals in detail. In order to validate the proposed method, a comprehensive data collection and experimental analysis have been carried out. \secref\ref{sec:exp_ana} \revxy{provides extensive evaluations and results}. Finally in \secref\ref{sec:conclusion}, we conclude the paper.

\section{Related work}
\label{sec:rel_work}
The main topic of this paper \revxy{involves} the information fusion using \rev{graph-based} optimization per crowd-sourcing in order to build \acrshort{ips} in a low-cost (both in time and labour) manner. We mainly present the literature related to methods for generating \acrshort{rfm} using \acrshort{rf} measurements. More details regarding feature-based positioning approaches and for extracting relative poses from \acrshort{imu} can be found \eg in \cite{zhou2019mitigating,he2016wifi} and \cite{yan2019ronin,liu2020tlio}, respectively.

\subsection{RFM construction approaches}
\textbf{Fully-supervised} \revxy{In} the early stage of building a feature-based \acrshort{ips}, \acrshort{rfm} was \revxy{created by} manually marking the reference locations and measuring the \acrshort{rf} measurements in stop-and-go manner\cite{bahl2000radar,youssef2005horus}. In \cite{zhou2019mitigating}, Zhou \etal introduced a high precision tracking system for obtaining the reference locations in order to collect data kinematically. These approaches yield highly accurate reference locations, however, they are time-costing and labour-intensive. These two drawbacks \revxy{greatly} hinder the \revxy{widespread} deployment of feature-based \acrshort{ips}.

\textbf{Weakly-supervised} Compared to fully-supervised approaches, weakly-supervised ones incorporate other information \revxy{to simplify} the data collection procedure. The authors in \cite{liu2018implementation,liu2017quick} illustrated a quick \acrshort{rfm} creation method, which requires the site surveyor (or the contributor \revxy{of} the crowd-sourcing) to put starting and ending points on the map of the indoor region of interest. The reference locations are interpolated by assuming that the surveyor is moving along a straight line. This can speed-up data collection but requires indoor maps as well as the explicit interaction from surveyors. Another way to reduce the labour for collecting the \acrshort{rfm} is by combining \rev{fully-supervised methods} with reconstruction algorithms. A sparse data collection is firstly performed in a fully-supervised way. The sparsely collected \acrshort{rfm} is used to reconstruct a dense \acrshort{rfm} either by spatial interpolation\cite{feng2011received,khalajmehrabadi2016joint}, \eg compressive sensing\cite{feng2011received}, Gaussian process regression\cite{atia2012dynamic} or by employing the path-loss model of the \acrshort{rf} signals\cite{belmonte2018radiosity}. However, the former \rev{is constrained} by the granularity of existing reference locations, and the latter requires to know the location of signal sources \rev{in prior} and to estimate the parameters of the propagation model for each signal source in different indoor regions.

\textbf{Unsupervised} There are three types of \rev{approaches to} marking the reference locations in an unsupervised manner: i) with the help of floor plan\cite{wu2012will,shen2013walkie}; ii) with the help of moving robots\cite{sen2012you,chen2018slide}; and iii) automatic trajectory fusion\cite{gu2017,gu2019,nowicki2015indoor,nowicki2019multi,tan2018optimization,zuo2018indoor}. In \cite{wu2012will} and \cite{shen2013walkie}, the reference locations are estimated by matching \acrshort{rf} measurements with geometric features (\eg connectivity \rev{between} rooms, corridors, and corners) of the floor plan. PinLoc\cite{sen2012you} and Slide\cite{chen2018slide} employed automatically controlled robots to improve the efficiency of site survey. However, it either requires indoor maps or \revxy{introduces} extra hardware. For the trajectory fusion approaches, they combined relative pose information, mostly from \acrshort{imu}, with constraints derived from \rev{measurable radio signals}. The reference locations are obtained via solving an optimization problem. These methods are efficient and \revxy{do} not require additional support. However, they need to model constraints between \acrshort{rf} measurements.

\rev{Our proposal is in an unsupervised manner. In contrast to previous work, our approach does not require any known reference locations as landmarks nor the help of dedicated surveying devices and detailed indoor maps.}

\subsection{Geometric constraints generation}
When \revxy{using} \rev{graph-based optimization} approaches \revxy{to fuse} the trajectories, the key is \revxy{to build} data associations. In \cite{gu2017,gu2019}, Gu \etal proposed to build the \acrshort{rfm} in a crowd-sourced way by aligning multi-trajectories using \rev{graph-based} optimization. But it deployed the foot-mounted \acrshort{imu}, which can yield accurate relative pose estimations. Nowicki \etal \cite{nowicki2015indoor,nowicki2019multi} and Tan \etal\cite{tan2018optimization} also employed graph optimization for aligning trajectories but it relies on known locations as landmarks \revxy{to build} the spatial relationship. These \revxy{works} employ a quasi-linear model to approximate the spatial constraints between trajectories using radio signals and cannot estimate the uncertainty of the geometric constraints. In addition, it requires \rev{priorly measured} location of landmarks. In \cite{zuo2018indoor} authors employed the path-loss model for estimating the spatial relationship between relative locations. The main challenge of this approach is that parameters of the propagation model \revxy{vary} for each signal source as well as indoor environments. In \cite{liu2019collaborative}\rev{, an approach based on similarity threshold between radio signals is used to detect data association. It utilizes the relative pose information to estimate the confidence (\ie level of uncertainty) of data associations. However, it cannot model the expected distance between nodes but simply treats them at the same position if the similarity between radio signals is higher than the threshold.} \revxy{Compared} to previous work, our proposal is in a self-supervised manner, namely modelling the geometric information (both the expected distance and its level of uncertainty) adaptively using the relative locations and their associated radio signals.

\section{Information fusion as graph optimization}
\label{sec:problem}
\subsection{Representations}
For one trajectory collected by crowd-sourcing contributors, it consists data from \acrshort{imu} and \acrshort{rf} measurements. Each of the trajectory is used to extract relative spatial relationship according to \acrshort{imu} data, \eg per traditional \acrfull{pdr}\cite{hostettler2016imu} or learning-based approaches\cite{liu2020tlio,yan2019ronin}. Thus, the $i$-th trajectory $ \mathbf{T}_i $ is represented using a sequence of relative positions and their associated observations, $ \mathbf{T}_i = \{(x_i^{(k)}, y_i^{(k)}, \theta_i^{(k)}, \mathbf{O}_i^{(k)})\} $, where $ x_i^{(k)}, y_i^{(k)} , \theta_i^{(k)}$ is the 2D pose (position plus heading) of the $ k $-th location in a local coordinate frame of the trajectory, and $ \mathbf{O}_i^{(k)} $ is the $ k $-th observation associated to the corresponding location in case of opportunistic \acrshort{rf} signals are measured. Without \revxy{loss of} generality, \acrshort{rf} measurements from Wi-Fi/\acrshort{ble} beacons are taken as examples. Each \acrshort{rf} observation consists a set of paired values, \ie a unique identifier (e.g., MAC) and its corresponding value (e.g., received signal strength indicator or channel state information). Thus, the $ k $-th observation $ \mathbf{O}_i^{(k)} $ of $ i $-th trajectory $ \mathbf{T}_i $ is denoted as $ \mathbf{O}_i^{(k)} := \{a: v | \mathbf{O}_i^{(k)}(a) = v, a \in \mathcal{A}_i^{(k)}\}$, where $ \mathcal{A}_i^{(k)} $ is a set of identifiers of measurable signals.


In this paper, we represent the \rev{trajectory alignment as a graph-based optimization problem}, in which the node $ \mathbf{p}:= (x, y, \theta) $ is the 2D pose of the contributors, and the edge $ e_{ij}(\mathbf{p}_i, \mathbf{p}_j|\mathbf{O}_i, \mathbf{O}_j):=e_{ij} $ used to link these nodes is either from \acrshort{imu} measurements \rev{(temporal constraints)} or from radio signals associated with each node \rev{(geometric constraints)}. The goal of trajectory alignment is to \rev{estimate the transformation matrix between associated nodes in order to transform all trajectories into a common coordinate frame.} The approach \revxy{to} achieving such a goal is via minimizing a defined loss function associated with all edges.


\subsection{Energy function and its minimization}
An observation model $ h $, denoting the function that builds edges between observations, is used to model expected observations \rev{for updating the nodes}, $ \mathbf{z}=h(\mathbf{p}) + \epsilon $, where $ \epsilon \in \mathcal{N}(0, \Omega^{-1}) $. The gap between modelled observations and expected ones denotes the error term that needs be minimized via the optimization. Following \cite{grisetti2011g2o}, the error function defined as negative log-likelihood of Gaussian priors can be formulated as a \acrfull{nls} problem:
\begin{equation}
	\label{eq:energy_fun}
	\hat{\mathbf{P}} = \underset{\mathbf{p}}{\operatorname{argmin}}(\sum_i\|h(\mathbf{p}_i) - \mathbf{z}_i\|_{\Omega_i}^{2})
\end{equation}
The above \acrshort{nls} problem can be solved efficiently using well-established solvers, e.g., Gauss-Newton, or Levenberg-Marquardt per gradient descent and it can be carried out via open-source libraries, such $ \mathrm{g}^2{\mathrm{o}} $ and Ceres. More details regarding \rev{graph-based} optimization can be found \eg in \cite{gu2019,nowicki2019multi,grisetti2011g2o}.

\subsection{Temporal spatial association from IMU measurements}
As presented in above section, the key to formulating the information fusion problem as \rev{graph-based} optimization is to build associations between nodes. In this paper, we mainly take two types of edges into account: i) \rev{(local) temporal} constraints between sequential positions obtained using \acrshort{imu} data; and ii) \rev{(global) geometric} constraints between arbitrary positions from the \acrshort{rf} measurements. The former type only exists intra-trajectory \rev{between consecutive nodes} and the latter can be establish in both intra-/inter-trajectories. In order to build each type of edges, an observation function has to \revxy{be provided} for computing the expected measurements and its uncertainty.

For the local spatial association, we follow \cite{gu2017,gu2019} and formulate it as motion control input from \acrfull{ins}. The observation model $ h^{\mathrm{IMU}} $ extracts the motion control input $ \mathbf{z}_{i+1}^{\mathrm{IMU}} $ from a given pair of sequential nodes, $ \mathbf{p}_i $ and $ \mathbf{p}_{i+1} $, within one trajectory, i.e. $ \mathbf{z}_{i+1}^{\mathrm{IMU}} = h^{\mathrm{IMU}}(\mathbf{p}_i, \mathbf{p}_{i+1}) + \epsilon_{i, i+1}^{\mathrm{IMU}}$, where
\begin{equation}
	\label{eq:h_imu}
	h^{\mathrm{IMU}}(\mathbf{p}_i, \mathbf{p}_{i+1}) = 
	\begin{bmatrix}
	\operatorname{cos}(\theta_i) & - \operatorname{sin}(\theta_i) & 0 \\
	\operatorname{sin}(\theta_i) & \operatorname{cos}(\theta_i) & 0 \\
	0 & 0 & 1 
	\end{bmatrix}
	\Delta \mathbf{p}_{i, i+1}	
\end{equation} 

\begin{equation}
	\Delta \mathbf{p}_{i, i+1}=
	\begin{bmatrix}
		x_{i} - x_{i+1}\\
		y_{i} - y_{i+1}\\
		\theta_{i} - \theta_{i+1}
	\end{bmatrix}
\end{equation}
and $ \epsilon_{i, i+1}^{\mathrm{IMU}} \in \mathcal{N}(0, (\Omega_{i, i+1}^{\mathrm{IMU}})^{-1}) $, where
\begin{equation}
	\label{eq:ep_imu}
	\Omega_{i, i+1}^{\mathrm{IMU}} =
	\begin{bmatrix}
	a & 0 & 0\\
	0 & a & 0\\
	0 & 0 & b
	\end{bmatrix}.
\end{equation}

The expected control input is obtained according to an \acrshort{ins} calculated using \acrshort{imu} measurements and current control input is computed per optimized value of nodes. The information matrix in \eqref{eq:ep_imu} is related to performance of the \acrshort{ins}. In this paper, we adopt the state-of-the-art learning-based approach, \textit{RoNIN}\cite{yan2019ronin}, for inferring the relative position. \rev{The value of coefficients $ a $ and $ b $ are set to} 500 and 70, respectively. In this way, the information matrix does not take the diversity of devices and \revxy{motions} into account. Diversity consideration is relevant \revxy{to} improving the generality and is left for the future work.

\subsection{System framework}
\begin{figure}
	\centering
	\includegraphics[width=0.5\columnwidth]{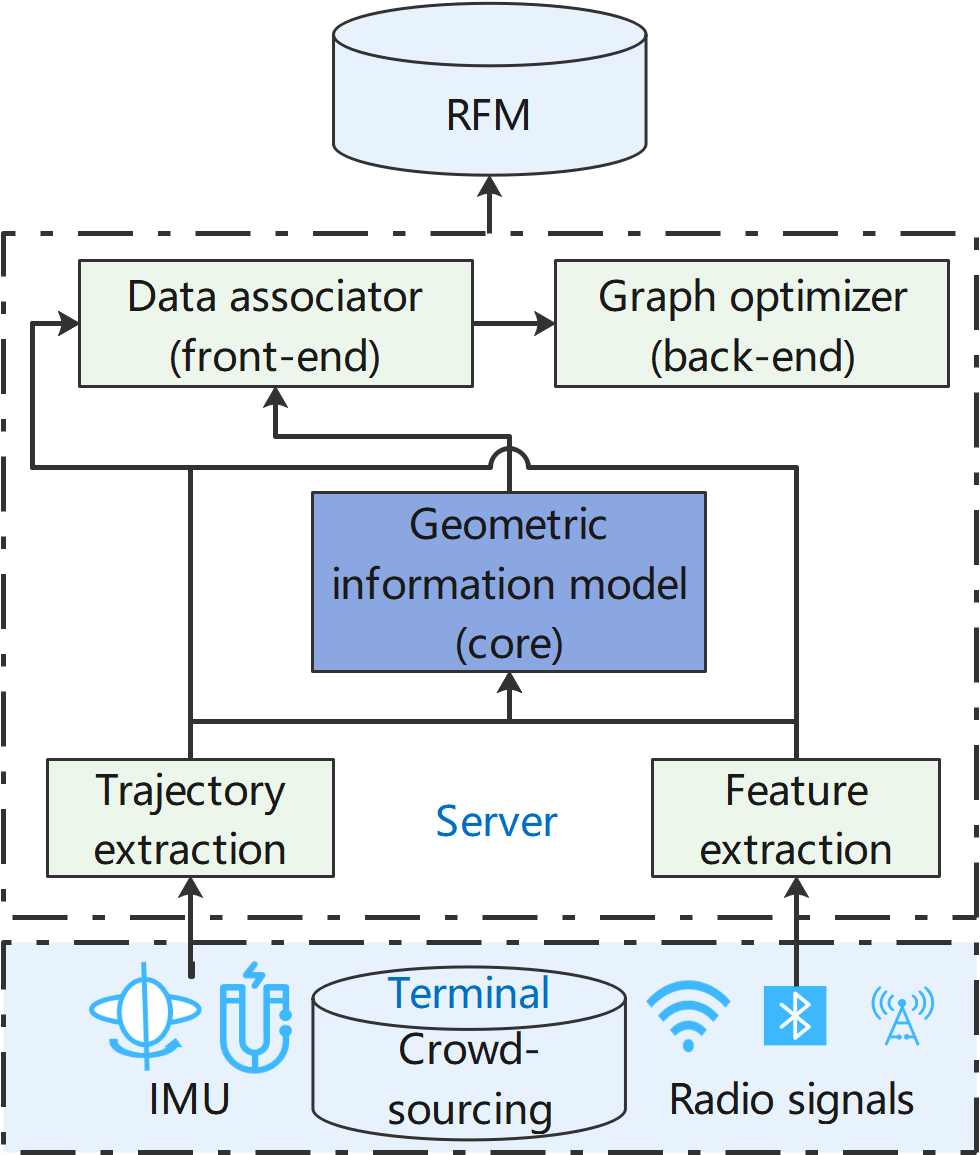}
	\caption{\rev{The scheme of proposed framework}}
	\label{fig:sys_scheme}
\end{figure}
\figref\ref{fig:sys_scheme} \rev{illustrates the proposed solution for generating RFM using crowd-sourced trajectories. It starts with crowd-sourcing raw measurements, including IMU readings and radio signals, from various mobile devices. These raw data are processed on a server. Relative poses are extracted per learning-based dead reckoning approach using IMU measurements. Their corresponding radio signals are associated according to the temporal information. The core module is the geometric information modelling (presented in} \secref\ref{sec:spa_rf}\rev{), which takes extracted relative positions and radio signals as input and approximate geometric constraints according to similarity between radio signals. The adaptively estimated model is then used to build loop closures between trajectories in order to form a properly connected graph. The RFM is obtained per optimizing the graph using well-established graph optimizers.}
 
\section{Spatial association construction using radio signals}
\label{sec:spa_rf}
\begin{figure*}
	\centering
	\subfloat[Example from test area 1]{\includegraphics[width=0.45\linewidth]{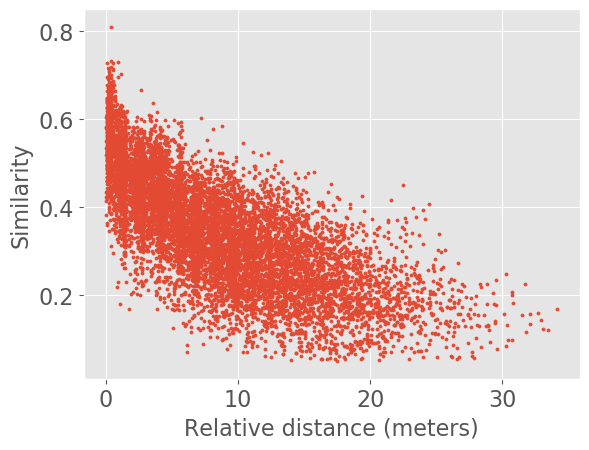}
		\label{subfig:dyc_sim_dist_f2}}
	\subfloat[Example from the test area 2]{\includegraphics[width=0.45\linewidth]{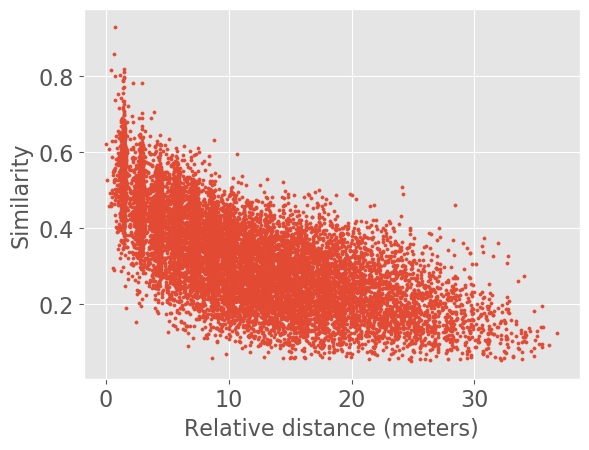}
		\label{subfig:wxtd_sim_dust_f1}}
	\caption{Examples of the relationship between similarity and distance}
	\label{fig:example_sim_dist}
\end{figure*}
Regarding the geometric constraints, they are constructed using \rev{RF measurements based on the fundamental assumption that different locations can be identified by measurable radio signals. However,} mining global spatial associations from radio signals is difficult due to three factors: i) although the propagation of \acrshort{rf} signals in free-space follows the log-path loss model, it is difficult to estimate the attenuation factor of each signal source in the indoor environment; ii) it requires the absolute location information \revxy{to estimate} the propagation model but it is not available in crowd-sourced trajectories; and iii) the estimation of the covariance matrix of \acrshort{rf} measurements is complicated because the \acrshort{rf} feature is in hyper-dimension.

Our proposal tackles the \acrshort{rf} measurement-based association construction problem according \revxy{to} the findings that a proper metric $ g $ between radio signals can reflect the distance in geometric space. We downgrade the problem of estimating the observation model for each \acrshort{rf} signal source at each location into \rev{a} problem of approximating the spatial constraint according to the metric between each pair of \acrshort{rf} measurements. In this way, it can make full use of the distance between relative positions inferred from \acrshort{imu} data.

Given a batch of crowd-sourced trajectories $ \{\mathbf{T}_i\}_{i=1}^{M} $, a fix number of paired relative positions, which have the associated \acrshort{rf} measurements, are arbitrarily sampled from each trajectory. It yields a set of paired positions and \acrshort{rf} measurements $ \mathbf{D}:= \{(\mathbf{p}_i^{(l)}, \mathbf{p}_j^{(l)}, \mathbf{O}_i^{(l)}, \mathbf{O}_j^{(l)})\}_{l=1}^{N} $. These pairs are used to approximate the observation model $ h^{\mathrm{RF}} $ \revxy{to build} the spatial constraints based on \acrshort{rf} measurements such that:
\begin{equation}
	\label{eq:h_rf}
	z_{ij}^{\mathrm{RF}} = h^{\mathrm{RF}}(\mathbf{p}_i, \mathbf{p}_j, \mathbf{O}_i, \mathbf{O}_j|d, g) + \epsilon_{i, j}^{\mathrm{RF}}
\end{equation}
where $ \epsilon_{i, j}^{\mathrm{RF}}\in\mathcal{N}(0, (\sigma_{ij}^{\mathrm{RF}})^{-1}) $. $ d $ and $ g $ are \rev{distance metrics} between relative positions and similarity measure between \acrshort{rf} measurements, respectively. \rev{Since to estimate the geometric information model using radio signals only requires information embodied in crowd-sourced trajectories, we can build geometric constraints in a fully-unsupervised (self-supervised) way.} As the global spatial relationship is \revxy{approximated} according to the similarity measure, only the expected distance and its \revxy{level of uncertainty} is quantified. Namely, the spatial constrained between a pair of relative positions is defined on a circle in 2D-space (or a sphere in 3D-space). The proposed approach consists following key steps (summarized in \tabref\ref{tab:app}):
\begin{figure}
	\centering
	\includegraphics[width=0.5\columnwidth]{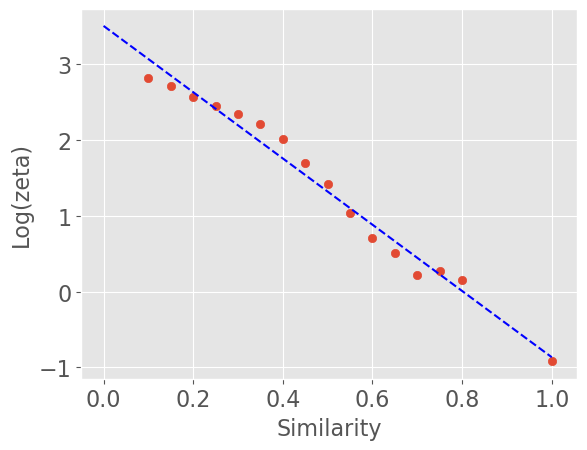}
	\caption{Example of log-linear relationship between $\zeta$ and similarity}
	\label{fig:log_linear_zeta_sim}
\end{figure}

\begin{itemize}
	\item \textbf{Metric computing}: for each data point in $ \mathbf{D} $, the distance between relative position and similarity measure are computed using the given $ d $ and $ g $. In this paper, $ d $ is defined as Euclidean distance between x-y coordinates of $ \mathbf{p}_i $ and $ \mathbf{p}_j $. It represents the geometric distance by assuming the local consistency of the coordinate frame within one trajectory. The corresponding similarity between \acrshort{rf} measurements is computed using a compound similarity measure, motivated by \cite{zhou2018cdm}, which combines Jaccard similarity\cite{zhou2019modified} and kernelized $ \mathcal{L}_1 $ similarity via harmonic mean ($ \beta $-score):
	\begin{equation}
		\label{eq:harmonic_mean}
		g(\mathbf{O}_i, \mathbf{O}_j) = \frac{(1+\beta^2)\cdot g^{\mathrm{Jac}}(\cdot, \cdot)\cdot g^{\mathrm{L1}}(\cdot, \cdot)}{\beta^2g^{\mathrm{Jac}}(\cdot, \cdot)+g^{\mathrm{L1}}(\cdot, \cdot)}
	\end{equation}
	where $ g^{\mathrm{Jac}}(\cdot, \cdot) $ denotes the Jaccard similarity:
	\begin{equation}
		\label{eq:jac}
		g^{\mathrm{Jac}}(\mathbf{O}_i, \mathbf{O}_j) = \frac{|\mathcal{A}_i\cap\mathcal{A}_j|}{|\mathcal{A}_i\cup\mathcal{A}_j|},
	\end{equation}
	and $ g^{\mathrm{L1}}(\cdot, \cdot) $ is defined as mapping the average of $ \mathcal{L}_1 $ distance to a similarity measure using Gaussian kernel function:
	\begin{equation}
		\label{eq:l1}
		g^{\mathrm{L1}}(\mathbf{O}_i, \mathbf{O}_j) =\operatorname{exp}(-(\frac{\sum_{a\in\mathcal{A}_{ij}}(|v_i - v_j|)}{|\mathcal{A}_{ij}|})^2/(2\sigma^2)).
	\end{equation}
	$ \mathcal{A}_{ij}:= \mathcal{A}_i\cup\mathcal{A}_j$. In case of $ a\notin\mathcal{A}_{i} $ or $ a\notin\mathcal{A}_{j} $, its corresponding value is set to a missing indicator. As both of the measured value of radio signals and estimated relative locations are affected by measurement noise, the relationship between distance and similarity \revxy{is} highly scattered as illustrated in \figref\ref{fig:example_sim_dist}.
	\begin{figure}
		\centering
		\includegraphics[width=0.5\columnwidth]{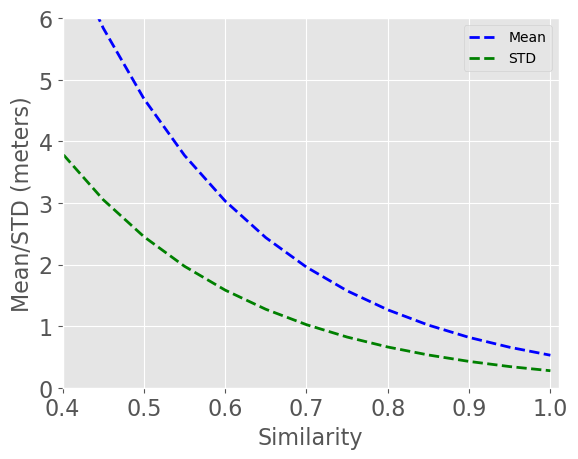}
		\caption{Example of fitted model of geometric constraints}
		\label{fig:mu_std_sim}
	\end{figure}
	\item \textbf{Similarity binning}: According to \eqref{eq:harmonic_mean}, the similarity is in $ [0, 1] $. In order to approximate the distribution of the geometric distance for a given similarity, a binning step is applied to divide the whole range into several bins with equal size. Each bin is denoted as $ [c_i - \Delta,c_i + \Delta ) $. The geometric distance whose corresponding similarity is within a given bin is assigned into the corresponding bin. \rev{Distances between relative positions} lying in a certain bin are used to approximate the distribution.
	\item \textbf{Distribution approximating}: For each bin, we estimate the expected geometric distance and its uncertainty (i.e. standard deviation). We approximate its distribution under Rayleigh distribution assumption parametrized with $ \zeta $ \cite{roy2004discrete}. Under this assumption, the approximated expected geometric distance and its variance is defined as:
	\begin{equation}
		\label{eq:ray_mu}
		\begin{split}
		&\mu_d = \zeta\sqrt{\frac{\pi}{2}}\\
		&\sigma_d^2 = \frac{4-\pi}{2}\cdot\zeta^2
		\end{split}
	\end{equation}
	where $ \zeta $ is the parameter for the Rayleigh distribution of the geometric distances in a given bin. The value of $ \zeta $ is estimated via a density estimator (e.g., kernel density estimation\cite{kim2012robust}) owing to the fact that the maximum value of probability is achieved at the value of $ \zeta $ when assumed as the Rayleigh distribution or maximum likelihood estimator\cite{roy2004discrete}.
	\item\textbf{Parameter estimating}: Assuming that the similarity range is divided into $ C $ bins, we can have a set of paired the similarity and its distribution parameter $ \{c_i, \hat{\zeta}_i\}_{i=1}^{C} $. From which, we would like to estimate a parametric model $ f $ such that given a similarity value $ g(\mathbf{O}_i, \mathbf{O}_j) $, it maps to an estimated $ \zeta $ value, i.e. $ f:g(\mathbf{O}_i, \mathbf{O}_j)\mapsto\zeta $. Thus, the expected geometric distance and its variance can be calculated using \eqref{eq:ray_mu}. This parametric model is assumed as log-linear for avoiding negative approximation of $ \zeta $ as well as the linearity between logarithm of $ \zeta $ and the similarity as shown in \figref\ref{fig:log_linear_zeta_sim}. The parameters are obtained per linear fitting. These estimated parameters can be applied to infer geometric constraints in a continuous space (see \figref\ref{fig:mu_std_sim}).
\end{itemize}

\renewcommand{\arraystretch}{1.25}
\begin{table}
	\caption{Summary of the proposed approach}
	\label{tab:app}
	\centering
	\begin{tabular}{p{0.06\columnwidth}p{0.9\columnwidth}}
		\hline
		\textbf{Inputs} & A batch of trajectories $ \{\mathbf{T}_i\}_{i=1}^{M} $\\
		&Geometric distance metric $ d $\\
		&Similarity measure $ g $\\
		\textbf{Output} & $ f: g(\mathbf{O}_i, \mathbf{O}_j)\mapsto\zeta $\\
		\hline
		1: & Sampling paired relative positions with \acrshort{rf} measurements from each trajectory\\
		2: & Computing the geometric distance and similarity measure for each pair of data points using the given $ d $ and $ g $\\
		3: & Binning the similarity range into $ C $ bins and assigning the geometric distance to the corresponding bins\\
		4: & Estimating the parameter $ \hat{\zeta}_i $ of the assumed Rayleigh distribution for each bin centred at similarity value $ c_i $ using kernel density estimators or maximum likelihood estimation\\
		5: & Estimating the parameter of $ f $ per linear fitting under log-linear assumption using $ \{c_i, \hat{\zeta}_i\}_{i=1}^{C} $\\
		\hline 
	\end{tabular}
\end{table}

\begin{figure}[!t]
	\centering
	\subfloat[Floor plan of the test area 1]{\includegraphics[width=0.5\columnwidth]{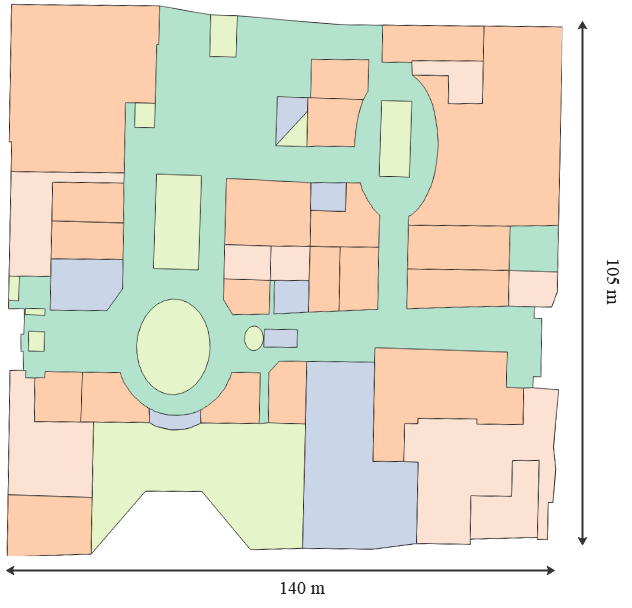}
		\label{subfig:dyc_f1}}\\
	\subfloat[Floor plan of the test area 2]{\includegraphics[width=0.5\columnwidth]{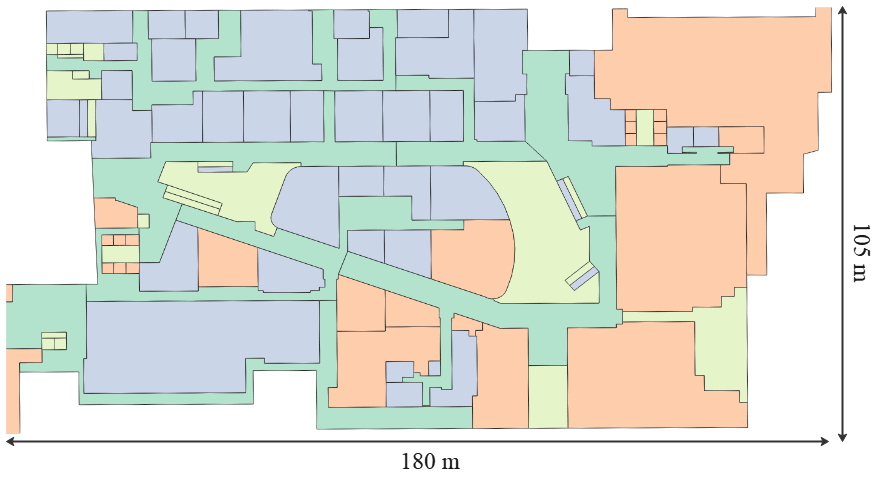}
		\label{subfig:wxtd_f1}}
	\caption{Floor plan of test areas (Green parts \rev{indicate} the accessible areas.)}
	\label{fig:test_areas}
\end{figure}

\section{Experimental analysis}
\label{sec:exp_ana}
\subsection{Experimental setup}
\begin{figure} 
	\centering
	\subfloat[Test area 1 (F1)]{\includegraphics[width=0.28\linewidth]{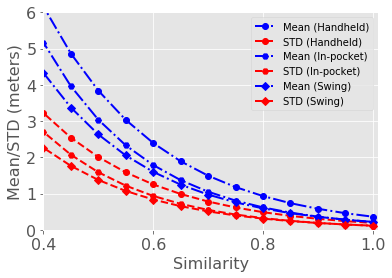}}
	\hspace{1ex}
	\subfloat[Test area 1 (F3)]{\includegraphics[width=0.28\linewidth]{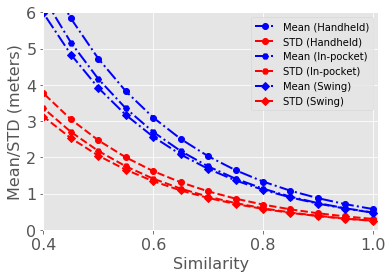}}
	\hspace{1ex}
	\subfloat[Test area 1 (F5)]{\includegraphics[width=0.28\linewidth]{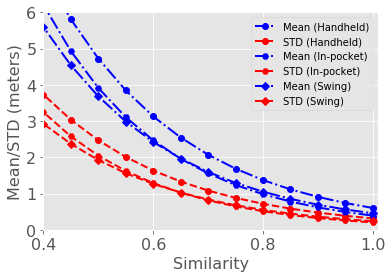}}
	\hspace{1ex}
	\caption{Impact of walking poses on geometric information modelling}
	\label{fig:cmp_fpm_poses}
\end{figure}
\subsubsection{Dataset}
In order to evaluate the performance of proposed approach, the crowd-sourcing procedures \rev{have been carried out } in two shopping malls illustrated in \figref\ref{fig:test_areas}. Each of them consists multi-storey and each floor covers an area over 10,000~$ \mathrm{m}^2 $. \revxy{The data were collected by} a custom-developed application \revxy{which can obtain} readings from all built-in sensors of the mobile device. \rev{Without losing the generality, only radio signals from WiFi access points (BLE beacons are omitted but includes mobile hotspots) are used for building the data associations as well as for indoor positioning.}

The crowd-sourcing is carried out using various type of mobile devices \footnote{\rev{The crowd-sourcing is carried by recruiting beta-test participants among all Huawei mobile service users under their agreement.}} by different volunteers without constraining walking poses \rev{(\eg static hold or swing)} and \revxy{attachments} of mobile devices \rev{(\eg handheld or in-pocket)}. \rev{We denote the data collection as commercialized-crowd-sourcing (CCS). Although it requires volunteers to report their walking poses and attachment of mobile devices as well as roughly mark the area where they have crowd-sourced (\eg the floor and passed-by accessible areas), these data are only used for the purpose of validation.} \revxy{Knowing the floor of each trajectory in advance is to reduce the influence of mis-identification of the traces' floor}. \rev{Automatically identification of the floor of trajectories (or a bunch of traces) is relevant for generating the RFM. However, it is out-of-scope of this paper and left for future work.}

\rev{For quantitative evaluation of the positioning accuracy, we manually collected the RFM (denoted as manually site surveying (MSS)) and test points} \revxy{in both} \rev{test areas using an unmanned ground vehicle, which is mounted with LiDAR and the mobile device. It is capable of performing visual-SLAM in centi-meter accuracy. The mapped test areas are used to obtain the reference locations and the ground truth of test points for evaluating the positioning performance. The detail information of collected data is summarized in }\tabref\ref{tab:sum_data} \footnote{\rev{The summary of APs takes all measured ones into account without e.g., identifying mobile APs or detecting changes of fixed APs.}}.

\setlength{\tabcolsep}{2pt}
\begin{table}
	\caption{\rev{Summary of collected data}}
	\label{tab:sum_data}
	\centering
	\begin{tabular}{c|c|cccc|cc|cc|c}
		\hline
		&\multirow{2}{*}{Floor}&\multicolumn{4}{c}{CCS}\vline&\multicolumn{2}{c}{MSS}\vline&\multicolumn{2}{c}{Test dataset}\vline&\multirow{2}{*}{Total APs}\\
		\cline{3-10}
		&&$\sharp$ Traj.&Length (m)&$\sharp$ Samples&$\sharp$APs&$\sharp$Samples&$\sharp$APs&$\sharp$Samples&$\sharp$APs&\\
		\hline
		\multirow{5}{*}{\shortstack{Test \\area 1}}&F1&89&1991&1149&1617&1098&2589&300&2112&2833\\
		&F2&115&2093&1136&1597&1218&2792&290&2156&2942\\
		&F3&135&2706&814&1578&807&2407&234&2159&2639\\
		&F4&137&2804&1357&1657&837&2435&290&2099&2673\\
		&F5&149&3502&1834&1637&821&2573&272&2043&2789\\
		\hline
		\hline
		\multirow{5}{*}{\shortstack{Test \\area 2}}&F1&121&2481&1698&1951&589&2363&341&1917&2958\\
		&F2&63&1327&850&1544&773&2619&219&1839&3009\\
		&F3&52&880&505&1274&517&1909&205&1454&2271\\
		&F4&57&1055&648&1150&518&1555&234&1369&1867\\
		&F5&61&1013&610&1025&662&1763&280&1253&2103\\
		\hline
	\end{tabular}
\end{table}

\subsubsection{Short on positioning approach}
After \revxy{using} \rev{graph-based} optimization \revxy{to align} the crowd-sourced trajectories, all relative locations can be mapped into a common coordinate frame. These locations and their associated radio signals are used as the \acrshort{rfm} for feature-based indoor positioning. Herein we employ $k$-nearest neighbour search \cite{bahl2000radar,zhou2019modified}, a widely used deterministic feature-based positioning method, \revxy{to evaluate} the \revxy{positioning} performance. The \rev{distance} metric used to match radio signals is \rev{defined} as $ 1- g(\mathbf{O}_i, \mathbf{O}_j) $ and \rev{the parameter $ k $ equals to 5 and 10 for test area 1 and 2} \rev{via grid search}, respectively.

\subsubsection{Evaluation metrics}
For the geometric information modelling, we mainly evaluate and compare the generality qualitatively by modelling geometric constraints using different data sources, which are obtained in various walking poses, placement of mobile devices and indoor regions. In addition, we also evaluate the trajectory alignment qualitatively via visualizing it with the floor plan of indoor regions. Regarding the positioning performance, Euclidean distance between the estimate position and the corresponding ground truth is computed. The \acrfull{ecdf} of positioning error and its statistical values (\eg \acrfull{cep}) are used as the quantitative evaluation metrics.

\subsection{Geometric constraint modelling}
\begin{figure}
	\centering
	\subfloat[Test area 1]{\includegraphics[width=0.5\columnwidth]{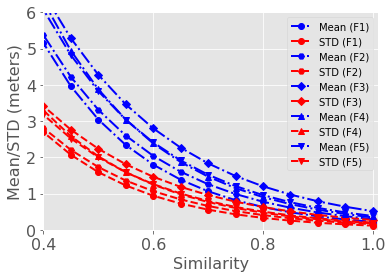}}
	\subfloat[Test area 2]{\includegraphics[width=0.5\columnwidth]{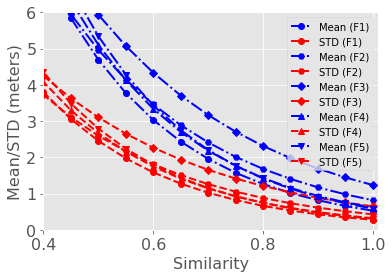}}
	\caption{Impact of indoor regions on geometric information modelling}
	\label{fig:cmp_fpm_rois}
\end{figure}

\begin{figure} 
	\centering
	\subfloat[Test area 2 (F1)]{\includegraphics[width=0.28\linewidth]{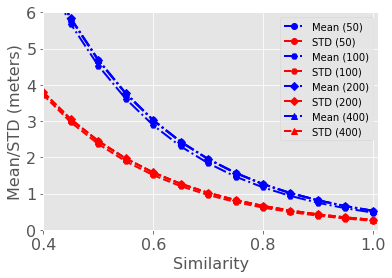}}
	\hspace{1ex}
	\subfloat[Test area 2 (F3)]{\includegraphics[width=0.28\linewidth]{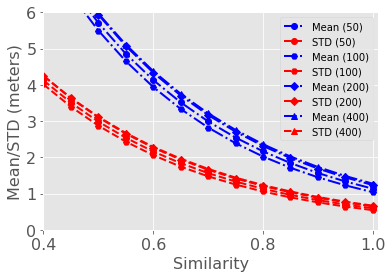}}
	\hspace{1ex}
	\subfloat[Test area 2 (F5)]{\includegraphics[width=0.28\linewidth]{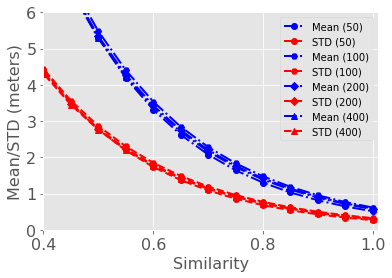}}
	\hspace{1ex}
	\caption{Impact of the number of sampled pairs on geometric information modelling}
	\label{fig:cmp_fpm_sampling_nums}
\end{figure}

In order to validate the proposed approach \revxy{to} modelling the geometric constraints, we evaluated and compared geometric information modelling results in case of various walking poses, indoor regions and the number of sampled pairs. As shown in \figref\ref{fig:cmp_fpm_poses} the impact of walking poses on the estimated geometric constraints regarding the given value of the similarity varies over different indoor regions, however the range of the variation is limited, especially when the similarity is higher than $ 0.6 $. Regarding the impact of indoor regions, the results illustrated in \figref\ref{fig:cmp_fpm_rois} \revxy{depict} that the consistency of geometric constraint models is high in most indoor regions over different floors and buildings. It means that the pre-estimated model can be transferred from one indoor region to another for improving the generalizability. Compare to the impact of walking poses and indoor regions on the geometric modelling, the sampling of the paired relative position and \acrshort{rf} measurement has much less influence as shown in \figref\ref{fig:cmp_fpm_sampling_nums}. \revxy{It} reveals the characteristic of the robustness of the proposed approach against arbitrary sampling. \rev{In the following experimental analysis, the geometric model estimation is carried out using crowd-sourced trajectories for each floor with diverse walking poses. The resampling number is set to 100.} 


\subsection{Trajectory fusion and indoor positioning}
\subsubsection{Qualitative evaluation of trajectory fusion}
\begin{figure}
	\centering
	\subfloat[0.25]{\includegraphics[width=0.28\linewidth]{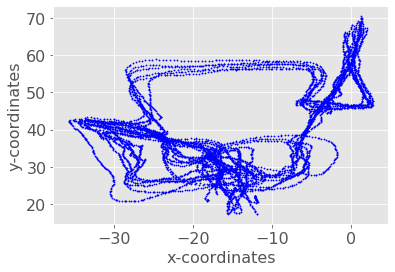}}
	\hspace{1ex}
	\subfloat[0.45]{\includegraphics[width=0.28\linewidth]{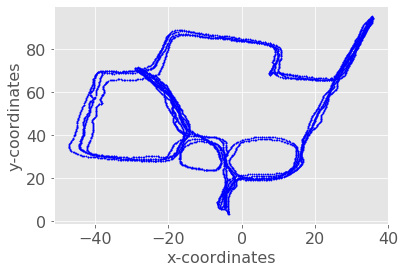}}
	\hspace{1ex}
	\subfloat[0.65]{\includegraphics[width=0.28\linewidth]{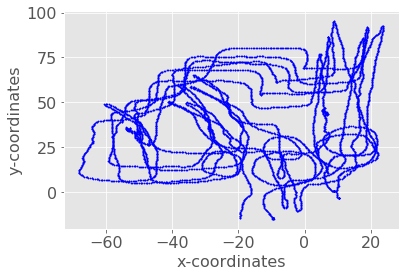}}\\
	\subfloat[0.25]{\includegraphics[width=0.28\linewidth]{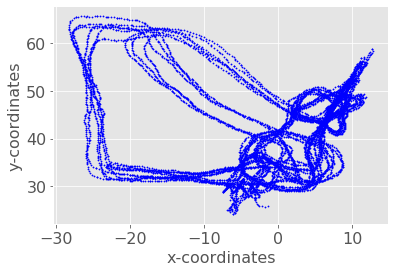}}
	\hspace{1ex}
	\subfloat[0.45]{\includegraphics[width=0.28\linewidth]{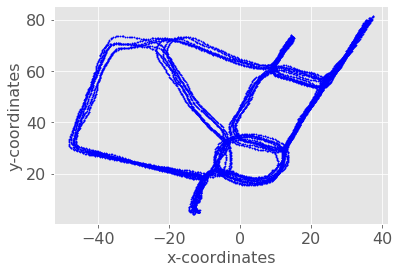}}
	\hspace{1ex}
	\subfloat[0.65]{\includegraphics[width=0.28\linewidth]{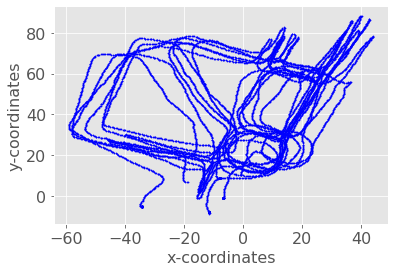}}\\
	\caption{\rev{The impact of the threshold on trajectory alignment (Top: F3 of test area 1/Bottom: F5 of test area 1)}}
	\label{fig:qua_aligned_dyc_thres}
\end{figure}

The trajectory fusion is achieved using graph optimization based on the open-source framework provided by $ \mathrm{g^2o} $ and the data associations between relative locations are built according to the estimated geometric information model for each indoor region\footnote{\url{https://github.com/RainerKuemmerle/g2o/tree/master/g2o}}. \rev{The key to aligning crowd-sourced trajectories is to build the spatial association between them. The number of generated edges between nodes is balanced by choosing the value of threshold on similarity between radio signals. Once data associations were established, graph-based optimization\footnote{\rev{In the implementation of $ \mathrm{g^2o} $, a built-in edge pruning based robust kernels is also used to remove the false connections.}} can be employed to align relative trajectories into a common reference frame (following the procedures presented in }\figref\ref{fig:sys_scheme}\rev{). The value of the threshold determines the connectivity between trajectories. The higher the threshold on similarity, the less the number of edges between trajectories. We threshold the similarity from 0.05 to 0.95 with interval of 0.1. As presented in }\figref\ref{fig:qua_aligned_dyc_thres}\rev{, we can obtain that there are too many false edges between trajectories, the crowd-trajectories are condensed too much when the threshold is small (e.g., 0.25). With a high threshold (e.g., 0.65), it results in low connectivity between trajectories, which cannot provide adequate geometric constraints for estimating the transformation matrix between them. Therefore, the crowd-sourced trajectories cannot be aligned properly. By performing the grid search, the suitable value of the threshold equals to 0.45 for both test areas.}

\begin{figure}
	\centering
	\subfloat[Raw trajectories]{\includegraphics[width=0.28\linewidth]{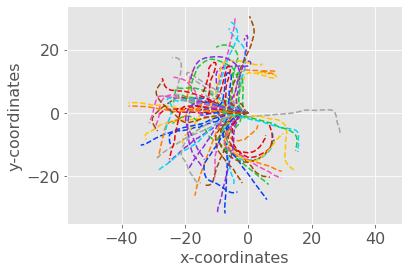}}
	\hspace{1ex}
	\subfloat[Aligned trajectories]{\includegraphics[width=0.28\linewidth]{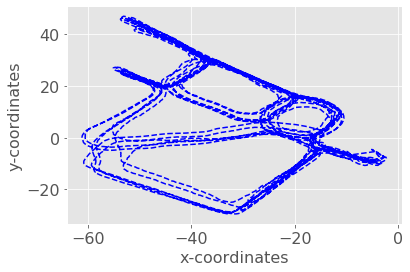}}
	\hspace{1ex}
	\subfloat[Aligned trajectories]{\includegraphics[width=0.28\linewidth]{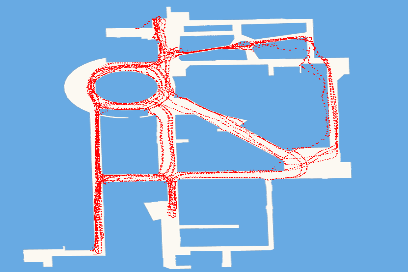}}\\
	\subfloat[Raw trajectories]{\includegraphics[width=0.28\linewidth]{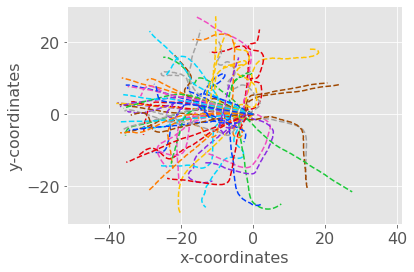}}
	\hspace{1ex}
	\subfloat[Aligned trajectories]{\includegraphics[width=0.28\linewidth]{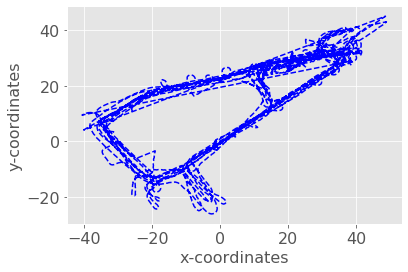}}
	\hspace{1ex}
	\subfloat[Aligned trajectories]{\includegraphics[width=0.28\linewidth]{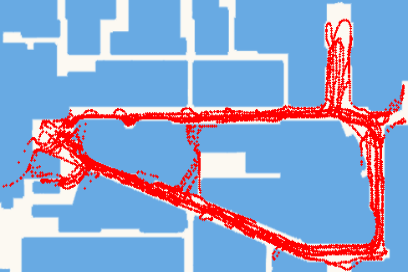}}\\
	\caption{Qualitative evaluation of aligned trajectories (Top: F5 of test area 1/Bottom: F1 of test area 2)}
	\label{fig:qua_aligned_wxtd}
\end{figure}

As shown in \figref\ref{fig:qua_aligned_wxtd}, the mis-aligned raw trajectories (1st column), which are in separate local reference frames, \revxy{seem} very noisy and without clear relations. After performing graph optimization according to the geometric constraints between different pairs of trajectories \rev{by thresholding on the similarity}, these raw trajectories can be fused together and aligned properly into one common reference frame (2nd column). Qualitatively, these aligned trajectories are well-matched to the floor plan of the indoor region per manual projection as illustrated in 3rd column of \figref\ref{fig:qua_aligned_wxtd}.

\subsubsection{Evaluation of positioning performance}
\begin{figure}
	\centering
	\subfloat[MSS]{\includegraphics[width=0.5\columnwidth]{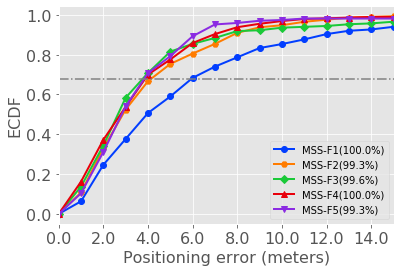}}
	\subfloat[CCS]{\includegraphics[width=0.5\columnwidth]{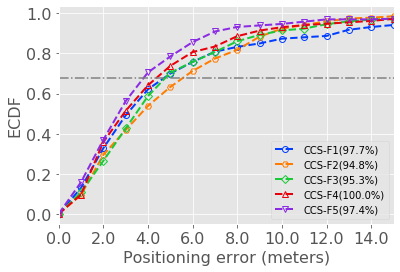}}\\
	\subfloat[MSS]{\includegraphics[width=0.5\columnwidth]{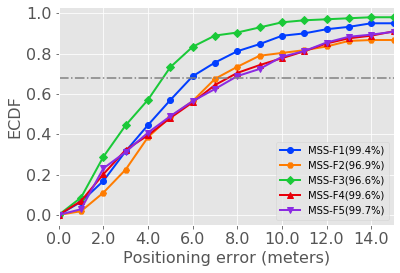}}
	\subfloat[CCS]{\includegraphics[width=0.5\columnwidth]{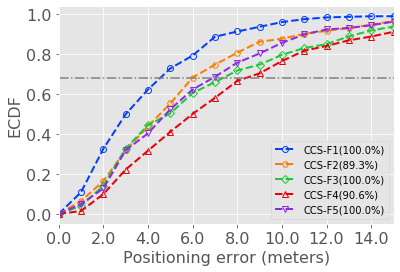}}
	\caption{\rev{ECDF of the positioning results (Top: Test area 1/Bottom: Test area 2). Number in the bracket denotes the floor identification accuracy. The dashed grey line in each figure indicates the CEP68.}}
	\label{fig:cmp_ecdf}
\end{figure}

Employing the \acrshort{rfm} generated per information fusion using the crowd-sourced trajectories \rev{or the RFM by site surveying}, the location of pedestrian can be determined by matching online measured radio signals with them. The statistical positioning results of two test areas are presented in \tabref\ref{tab:stats_pe} and the \acrshort{ecdf} of the positioning error is visualized in \figref\ref{fig:cmp_ecdf}. \rev{Regarding floor identification accuracy, the overall accuracy is about 97\% and 96\% using crowd-sourced RFM in both test areas. Compared to the floor detection accuracy using the RFM by MSS, the retreat is only 2 percentage point. For the location-wise positioning performance, the mean positioning error and CEP68 using crowd-sourced RFM are on a par with that using manually collected RFM in both test areas. The mean positioning error in test area 1 using CCS RFM is 4.3 meters, which falls back only 0.3 meters comparing to that of using MSS RFM. As in test area 2, the mean positioning error using CCS RFM is decreased by 0.4 meters when comparing using MSS RFM, whose mean positioning error is 6.4 meters.}

\setlength{\tabcolsep}{4pt}
\renewcommand{\arraystretch}{1.25}
\begin{table*}
	\caption{\rev{Statistical of positioning performance}}
	\label{tab:stats_pe}
	\centering
	\begin{tabular}{c|c|cc|cc|cc|cc|cc}
		\hline
		& & \multicolumn{2}{c}{\multirow{2}{*}{\shortstack{Floor\\Acc. (\%)}}}\vline & \multicolumn{8}{c}{Positioning error (m)}\\
		\cline{5-12}
		& & & & \multicolumn{2}{c}{Min.} \vline& \multicolumn{2}{c}{Mean} \vline& \multicolumn{2}{c}{CEP68} \vline& \multicolumn{2}{c}{CEP95}\\
		\cline{2-12}
		& & MSS & CCS&MSS & CCS&MSS & CCS&MSS & CCS&MSS & CCS\\
		\hline
		\multirow{6}{*}{\shortstack{Test \\area 1}} & F1 & 100&97.7& 0.2&0.1&5.4&4.8&6.0&4.7&15.5&15.3 \\
		& F2 & 99.3&94.8& 0.1&0.1&3.8&4.6&4.3&5.6&9.8&11.4 \\
		& F3 & 99.6&95.3& 0.1&0.2&3.8&4.6&3.9&4.9&12.8&12.0 \\
		& F4 & 100&100& 0.2&0.1&3.5&4.1&3.9&4.5&8.6&11.8 \\
		& F5 & 99.3&97.4& 0.2&0.1&3.6&3.6&3.8&3.8&6.9&10.0 \\
		\cline{2-12}
		& Overall & 99.6&97.0& 0.2&0.1&4.0&4.3&4.4&4.7&10.7&12.1 \\
		\hline
		\hline
		\multirow{6}{*}{\shortstack{Test \\area 2}} & F1 & 99.4&100& 0.2&0.1&5.6&3.9&6.0&4.5&14.0&9.5 \\
		& F2 & 96.9&89.3& 0.5&0.4&8.7&6.3&7.0&6.0&30.0&14.0 \\
		& F3 & 96.6&100& 0.3&0.4&4.1&6.3&4.7&7.3&9.7&15.2 \\
		& F4 & 99.6&90.6& 0.2&0.3&6.7&7.4&7.5&8.6&16.4&19.1 \\
		& F5 & 99.7&100& 0.5&0.2&7.0&6.0&7.9&6.9&18.2&14.2 \\
		\cline{2-12}
		& Overall & 98.4&96.0& 0.3&0.3&6.4&6.0&6.6&6.7&17.7&14.4 \\
		\hline
	\end{tabular}
\end{table*}

\section{Conclusion}
\label{sec:conclusion}
In this paper, we \revxy{propose} an approach for adaptively modelling the geometric information in a self-supervised manner using the crowd-sourced relative locations (\eg from inertial sensors or vision-inertial odometer) and their associated radio signals. The key to building the model is to \revxy{map} the similarity between radio signals to the physical distance under \rev{the assumption of Rayleigh distribution}. Geometric constraints obtained from radio signals can be used to fuse crowd-sourcing trajectories per solving an optimization problem. \revxy{Through} comprehensive experimental analysis, \revxy{we validate} the generalizability of the proposed approach and its applicability to trajectory alignment, which can generate the \acrshort{rfm} for feature-based indoor positioning. Using the \acrshort{rfm} generated from the crowd-sourced trajectories, \rev{the average floor identification accuracy is over 96\%, which is on a par with the one using manually collected RFM. The retreat is only about 2 percentage point. The mean positioning error is about 4.3 meters and 6.0 meters using crowd-sourced RFM in test area 1 and area 2, respectively. This is at the same level of using site surveyed RFM.} \rev{In order to alleviate the limitations of our proposal, the continuative research would focus on: i) the quantification of the uncertainty of relative positions obtained from IMU measurements; ii) the adaptiveness to long-term variations and device diversities; and iii) refinement of the crowd-sourced RFM.}
%
%

\section*{Acknowledgment}
The authors would like to thank the contribution of group members of positioning team at Riemann Lab, 2012 Laboratories, especially to Dr. Yun Zhang, Dr. Ran Guan, Xiya Cao, Jiaming Chen, and Mengchao Li, for their help of data processing and valuable discussions. Furthermore, we would like to thank the work of peer-reviews. Their suggestions have largely improve the quality of this paper.

\bibliographystyle{ieeetr}
\bibliography{./access_ref}

\end{document}